# Binary ZnSe:Fe$^{2+}$ and ternary ZnMgSe:Fe$^{2+}$ optical crystals for mid-IR applications

Sergei V. Naydenov[1*], Oleksii K. Kapustnyk[1], Igor M. Pritula[1],
Dmitro S. Sofronov[1], Igor S. Terzin[1], Nazar O. Kovalenko[1,2]

[1]Institute for Single Crystals of NAS of Ukraine, 60 Nauky ave., 61072 Kharkiv, Ukraine

[2]Helmut-Schmidt-Universität, Holstenhofweg 85, 22043 Hamburg, Germany

**Abstract.** In this study, binary ZnSe:Fe$^{2+}$ crystals and ternary Zn$_{1-x}$Mg$_x$Se:Fe$^{2+}$ crystals ($0 < x < 0.6$) were grown by the vertical Bridgman method in graphite crucibles under high argon pressure. A comparative characterization of the structural, energetic, and optical parameters of the obtained crystals was performed. Theoretical explanations of the observed experimental characteristics, including the specific redshift of the absorption and emission spectra with increasing solid-solution concentration, are provided. These results could be useful for the development of new laser media based on simple and complex AIIBVI crystals and for targeted engineering of their optical properties.

## Introduction

ZnSe-based crystals are among the most in-demand chalcogenide crystals. Undoped ZnSe crystals have long been used as optical windows, elements and substrates for infrared (IR) optics in the 1–20 µm range [1]. Scintillation crystals ZnSe activated by tellurium Te (more often) or aluminum are highly efficient for detecting X-rays or gamma rays with energies of up to several hundred keV [2].

The development of various optoelectronic and laser devices and systems that use or generate mid-infrared radiation in the most important atmospheric transparency range 2–5 µm requires the creation of functional laser materials that are efficient, compact, powerful enough, with a controlled and wide emission bandwidth, and resistant to adverse conditions and laser damage. In contrast to the visible range, there are few high-quality laser media for the mid-IR range. Oxygen atoms abnormally absorb radiation in this range. Therefore, the main role is transferred from oxygen-containing to chalcogenide laser materials. The most successful of these materials are binary AIIBVI crystals doped with isovalent ions of transition metals [3, 4].

A well-known medium for laser generation in the 2–3 µm range is the chromium-doped zinc selenide crystal, which was first used for this purpose in Livermore in 1996 [5]. Chromium-doped ZnSe:Cr$^{2+}$ crystals dominate the current market of infrared laser materials for the 2–3 µm emission range [3, 4]. The unique combination of physical, optical, and laser properties makes these crystals exclusive. However, owing to their practical needs (scientific, industrial, and military), extending their laser range to the 3–5 µm range is important. Possible ways to solve this problem are the use of iron Fe$^{2+}$ instead of chromium Cr$^{2+}$ as an activator, as well as the transition from binary to ternary crystals of solid solutions based on a chalcogenide matrix. To obtain coherent radiation in the range of 4–5 µm, active media based on ZnSe:Fe crystals doped with iron with a concentration of up to $10^{19}$ cm$^{-3}$ are often used [6, 7]. Experimental and theoretical studies [8-11] have confirmed that in Zn$_{1-x}$Mg$_x$Se:Cr$^{2+}$ and Zn$_{1-x}$Mg$_x$Se:Fe$^{2+}$ crystals, a strong "redshift" (long-wavelength shift) in the absorption/emission bands of several hundred nanometers can be obtained by increasing the solid solution concentration (without changing the type of crystal structure). This allows the laser emission band to shift to a more longwave region in the mid-IR range. On the other hand, flexible control of the emission bandwidth and wavelength tuning of active laser elements becomes possible by changing the activator concentration of the solid solution.

---

[*] sergei.naydenov@gmail.com

The aim of this study was to grow binary ZnSe:$Fe^{2+}$ and ternary $Zn_{1-x}Mg_xSe$:$Fe^{2+}$ ($0 < x < 0.6$) crystals doped with iron ions at approximately the same concentration, and to conduct a comprehensive investigation and comparative analysis of their structural, energetic, and optical parameters. Particular attention was paid to the influence of solid-solution concentration on the absorption and emission spectra of the activator ions, including the specific redshift effect observed for the absorption and emission bands with increasing solid-solution concentration.

**Experimental part**

Crystals of ZnSe:$Fe^{2+}$ and $Zn_{1-x}Mg_xSe$:$Fe^{2+}$ were grown by the high-pressure Bridgman method in a glassy carbon crucible under an external argon gas pressure 20-30 bar. The growth setup has a dedicated automated growth control system. This ensures a reduction in concentration fluctuations throughout the volume of a crystal ingot. The growth mode was adjusted by a temperature gradient of approximately 20–30°C/cm and a crucible travel speed of 1.2–1.7 mm/h. To obtain a solid solution crystalline $Zn_{1-x}Mg_xSe$:$Fe^{2+}$, a mixture of the binary compounds ZnSe and MgSe in the appropriate stoichiometric proportion was used. The raw material was thoroughly mixed and loaded into a graphite crucible. The crystals were subsequently grown according to the standard method of growing binary crystals of ZnSe:$Fe^{2+}$ by the Bridgman method under high pressure of an inert gas at an optimal melting temperature. We grew $Zn_{1-x}Mg_xSe$:$Fe^{2+}$ crystals throughout the whole range of solid solution concentrations between $x = 0.15$ and $x = 0.58$, in which this compound exists in the wurtzite (hexagonal) crystal structure. The ZnSe:$Fe^{2+}$ crystal corresponding to the composition $x = 0$ has a cubic crystal structure. The iron impurity concentration was on the order of several $10^{18}$ cm$^{-3}$. It consists of ~$10^{-3}$ wt.% of all the studied samples. All the ingots grown contained large single-crystalline blocks. The yield of the finished single crystals was approximately 50% relative to the total mass of the charged charge. We obtained a series of single-crystal ingots 30–50 mm in diameter and 50–100 mm in length (see Fig. 1). Optical elements of various shapes were removed and manufactured from the obtained crystals (see Fig. 2). Pairs of opposite (transmission) faces of optical elements were subjected to laser quality polishing. The sides of the optical elements were ground but not polished. Anti-reflective coatings were not used.

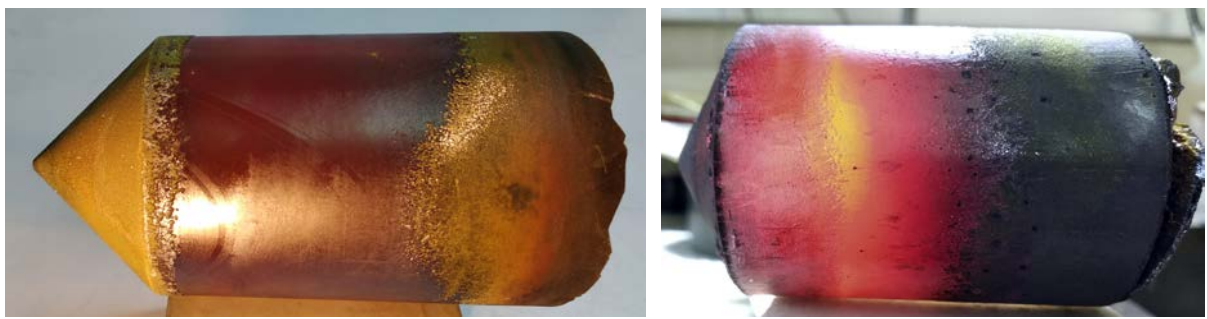

**Fig. 1.** Images of typical grown crystal ingots of ZnSe:$Fe^{2+}$ and $Zn_{1-x}Mg_xSe$:$Fe^{2+}$.

The elemental composition of the obtained crystals was determined using a JSM-6390LV scanning electron microscope (JEOL Ltd., Japan) equipped with an X-Max 50 X-ray energy-dispersive analysis detector (OXFORD Instruments Analytical, Great Britain).

The phase composition of the crystals was examined using an AERIS Research Edition powder X-ray diffractometer (Malvern PANalytical, Denmark).

The bandgap of the semiconductor crystals was determined by the optical absorption threshold method.

Absorption and emission spectra in the mid-IR, near-IR, and visible regions of the optical spectrum were studied.

Optical spectrophotometry was performed using a PerkinElmer Lambda 35 spectrophotometer (USA) in the range of 190–1100 nm and a PerkinElmer Spectrum One spectrophotometer (USA) in the range of 1.28–25 μm.

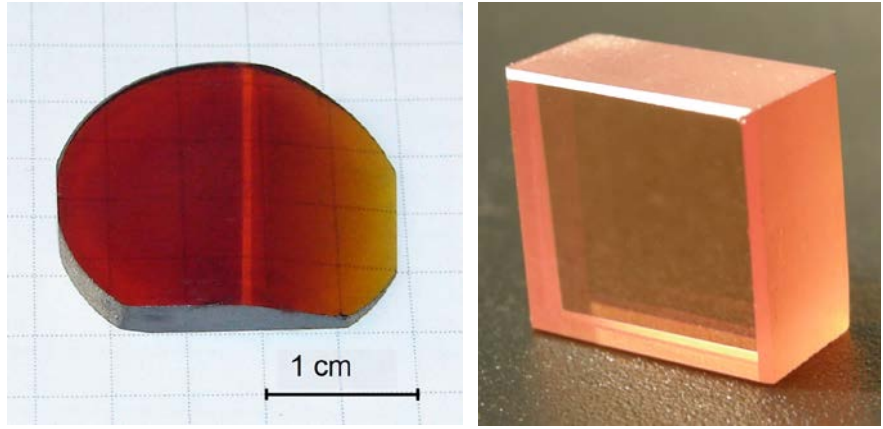

**Fig. 2.** Examples of optical elements (wafer, parallelepiped) cut and produced from the ZnSe:Fe$^{2+}$ and Zn$_{1-x}$Mg$_x$Se:Fe$^{2+}$ crystal ingots.

## Results and discussion

The ZnSe crystal has a zinc blende (sphalerite) structure with a cubic crystal lattice parameter of 5.66 Å [12]. Each central Zn ion is coordinated by four Se ions at the corners of the tetrahedron. The addition (cationic substitution) of Mg ions at a concentration of more than 6.5 at.% (corresponding to the solid solution concentration $x > 0.13$ for the Zn$_{1-x}$Mg$_x$Se crystals) leads to a change in the sphalerite structure to a wurtzite structure with hexagonal lattice parameters $a(x)$ and $c(x)$, which depend on the concentration $x$ of the solid solution. In the wurtzite structure, each Zn ion is also coordinated by four Se ions, which form a triangular pyramid. To characterize the changes in the environment of the Zn ions for both structures, two parameters are used: $d$ – the side of the base of the pyramid and $h$ – its height. They are related to the crystal lattice parameters by the relations $d_s = a/\sqrt{2}$ and $h_s = a/\sqrt{3}$ in the case of sphalerite ($s$) and $d_w = a$ and $h_w = c/2$ in the case of wurtzite ($w$). For some obtained crystals, the dependences of both unit cell parameters $a(x)$ and $c(x)$ on the different concentrations of Mg ions were measured by X-ray analysis (see Fig. 3). The Vegard law is not applicable with high accuracy since the change in these parameters significantly deviates from the linear trend at high solid solution concentrations. The change (difference) in the parameters of the coordination pyramid caused by a small increase in Mg concentration up to $x \approx 0.3$ is well described by a linear dependence $\Delta d(x) \approx k\Delta x$ or $\Delta c(x) \approx k\Delta x$, both of which have the same linear slope coefficient $k = 0.15$ Å.

Indeed, at a certain concentration $x_{s-w}$ of the Zn$_{1-x}$Mg$_x$Se solid solution, a transition from the sphalerite structure to the wurtzite structure occurs. Near this phase transition, the sphalerite tetrahedron transforms into a triangular wurtzite pyramid, i.e., the relation $h_s(x_{s-w}) = h_w(x_{s-w})$ is satisfied. From this, we can obtain the expression in the framework of the Vegard law approximation

$$k \approx \frac{\sqrt{3}}{2} c_{MgSe} - a_{ZnSe}. \tag{1}$$

Substituting the values of the lattice parameters $c = 6.72$ Å for MgSe (wurtzite) and $a = 5.66$ Å for ZnSe (sphalerite), we obtain $k = 0.16$ Å, which agrees well with the obtained experimental values.

We also studied the composition of the grown crystals. Coupled scanning electron microscopy with energy-dispersive X-ray spectroscopy and analysis of the composition of all obtained solid-solution crystals suggested good Zn/Mg substitution in the cation sublattice. The typical chemical element distribution along the ingot of Zn$_{1-x}$Mg$_x$Se:Fe$^{2+}$ is shown in Fig. 4. All the samples had satisfactory axial and radial homogeneity along and across the ingots. The magnesium content increases from the tip to the heel of the ingot since the Mg-impurity distribution coefficient in the ZnSe melt is significantly less than unity. When growing crystals of solid solutions of Zn$_{1-x}$Mg$_x$Se by the Bridgman method, eliminating this inhomogeneity is not easy, and this must be taken into account.

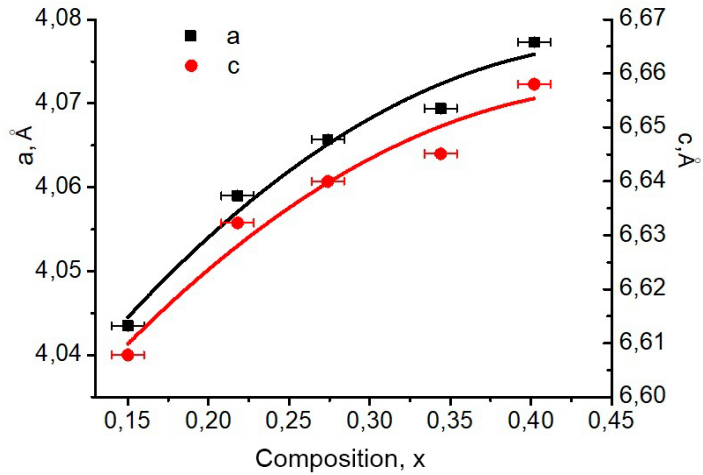

**Fig. 3.** The unit cell parameters *a* and *c* of $Zn_{1-x}Mg_xSe:Fe^{2+}$ crystals at different Mg concentrations.

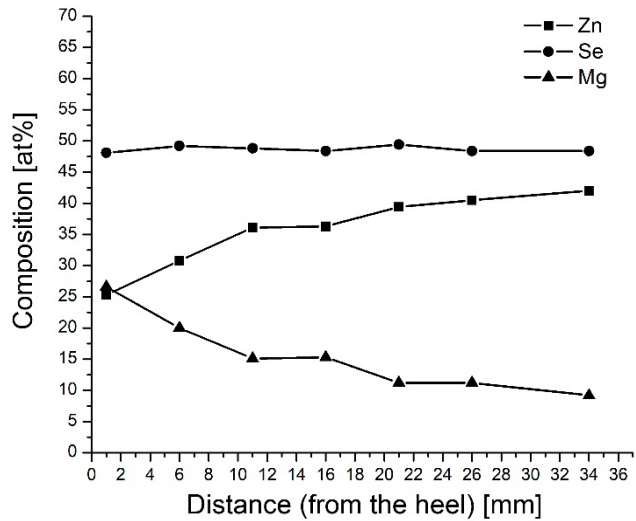

**Fig. 4.** Chemical element distribution of the solid solution of $Zn_{0.79}Mg_{0.31}Se:Fe^{2+}$ along the ingot.

The energy band-gap dependence of $Zn_{1-x}Mg_xSe:Fe^{2+}$ crystals on the solid solution composition was studied for different experimental samples with different concentrations from $x = 0.17$ (after the sphalerite-wurtzite phase transition) to $x = 0.58$ (see Fig. 5). The bandgap $E_g$ was determined via the optic absorption threshold method via the well-known relation $\alpha^2 \varepsilon^2 \approx B (\varepsilon - E_g)$ for direct-band semiconductors, where $\alpha$ is the experimental absorption coefficient, $B$ is a constant and $\varepsilon$ is the photon energy, which is valid for direct band–band optic transitions. As shown in Fig. 5, the dependence $E_g (x)$ has a linear character. The band gap increases with increasing solid solution concentration.

Notably, the dependence of the band gap change on the solid solution concentration remains linear over almost the entire range of $Zn_{1-x}Mg_xSe$ solid solution concentrations. In contrast, the dependences of the crystal lattice parameters on the solid solution concentration become nonlinear at sufficiently high solid solution concentrations. Usually, the opposite effect occurs in semiconductor crystals. The Vegard linear law can be fulfilled with good accuracy, whereas the dependence of the band gap on the composition change becomes nonlinear. We believe that the feature of $Zn_{1-x}Mg_xSe:Fe^{2+}$ crystals discovered here is associated with a strong difference in electronegativity for zinc (Pauling's electronegativity $\chi = 1.65$) and magnesium ($\chi = 1.31$). This affects the enhancement of the crystal field near the central ion of the transition metal (chromium or iron) surrounded by solid solution ligands. As a result, Stark splitting becomes stronger, and the redshift effect in these crystals becomes stronger (in more detail, see [10, 11]).

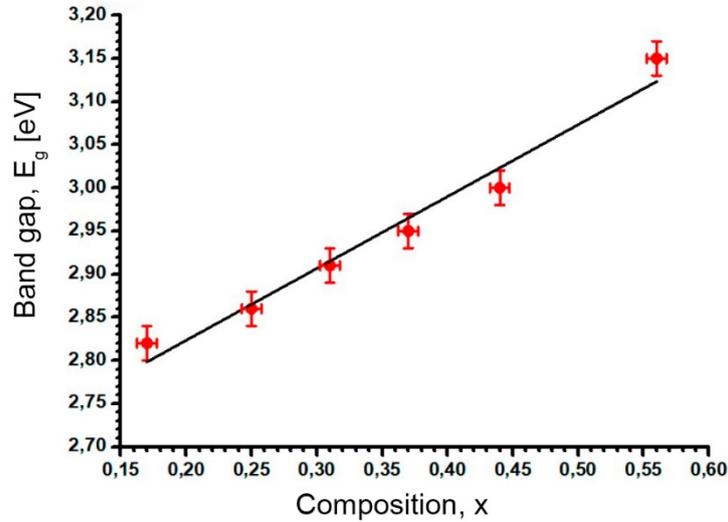

**Fig. 5.** Dependence of the energy band gap of $Zn_{1-x}Mg_xSe:Fe^{2+}$ crystals on the solid solution composition. The circles correspond to the data for different experimental examples.

The IR absorption and emission spectra of the grown optical crystals $ZnSe:Fe^{2+}$ and $Zn_{1-x}Mg_xSe:Fe^{2+}$ were studied. They are associated with intra-center transitions between the energy levels of the ground and excited states of the substitution ions $Fe^{2+}$, which are in the field of the surrounding ligands of the main matrix ZnSe. Fig. 6 shows the characteristic dependence of the bulk absorption coefficient for $ZnSe:Fe^{2+}$ crystals with a typical activator concentration of $\sim 5 \cdot 10^{18}$ cm$^{-3}$. Large values of the absorption coefficient correspond to the mid-IR absorption spectrum of these crystals. Taking into account the Jahn–Teller effect, the excited triplet state of the $Fe^{2+}$ ions in the energy band is split by the crystal field of the matrix into three sub-bands. Therefore, the absorption band can be represented as a combination of three Gaussian bands, which are shown in Fig. 6. In the wide wavelength region of 2.5–4.2 μm, strong absorption with a total maximum at 3.0–3.1 μm is observed.

The absorption spectra of the $Zn_{1-x}Mg_xSe:Fe^{2+}$ crystals with the same activator concentration of $\sim 5 \cdot 10^{18}$ cm$^{-3}$ have a similar appearance (see Fig. 7). Their important feature is the longwave redshift of the spectrum maximum with increasing concentration of the solid solution. The emission spectra of these crystals also revealed the indicated redshift effect. Fig. 8 show some experimental data and the theoretical linear dependence of the redshift effect. Analysis of these data reveals that the longwave shift in the absorption spectrum maximum in $Zn_{1-x}Mg_xSe:Fe^{2+}$ crystals is ~100 nm for every $\Delta x = 10\%$ change in the composition of the solid solution, and the shift in the emission spectrum maximum is ~80 nm/10% composition. This fairly significant shift in the spectra allows one to obtain laser radiation in the longwave wavelength range up to 5 μm.

The physical cause of the redshift effect in semiconductor solid-solution crystals is associated with the difference in energies ($\Delta E$) of the Stark splitting of the energy levels of transition-metal ions in the ligand field for the binary components of the solid solution [10]. This effect was previously discussed in detail for $Cd_{1-x}Mn_xTe:Fe^{2+}$ crystals [11]. In the present case, the splitting of $Fe^{2+}$ ions occurs within the $Zn_{1-x}Mg_xSe:Fe^{2+}$ matrix. The splitting energy of the iron ions in the $Zn_{1-x}Mg_xSe$ solid-solution matrix is determined according to the additivity principle, by analogy with Vegard's law for lattice parameters and/or the linear change in bandgap energy with solid-solution concentration. Therefore, with increasing solid-solution concentration $x$ (Mg content), the splitting energy of the $Fe^{2+}$ ion energy levels decreases correspondingly to the relation.

$$\Delta E(x) = \Delta E(ZnSe) - x\left[\Delta E(ZnSe) - \Delta E(MgSe)\right]. \tag{2}$$

The calculations revealed that the splitting energy in ZnSe matrix is greater than that in MgSe matrix, i.e., $\Delta E$(ZnSe) > $\Delta E$(MgSe). In other words, the crystal field acts on the iron ions in sphalerite ZnSe more strongly than in wurtzite MgSe. As a result, this leads to a redshift in the absorption and emission spectra to the long-wavelength region with increasing solid-solution concentration.

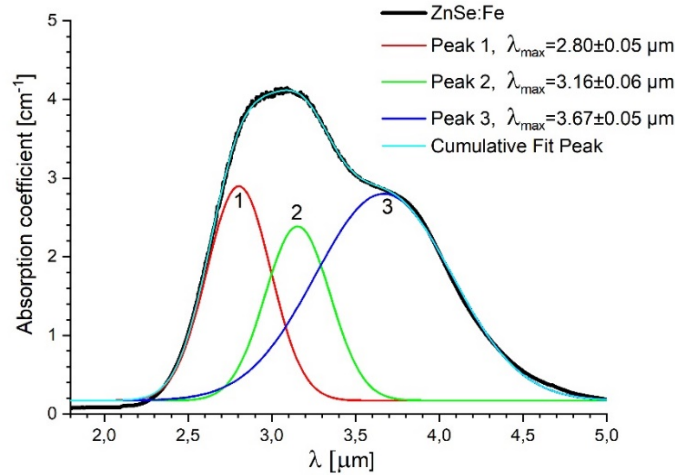

**Fig. 6.** Bulk absorption coefficient of the binary crystal ZnSe:$Fe^{2+}$ sample with a thickness of 5 mm (activator concentration ~$5 \cdot 10^{18}$ cm$^{-3}$).

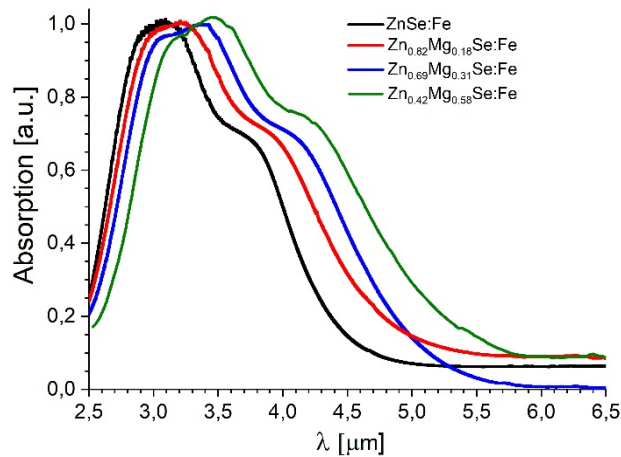

**Fig. 7.** Absorption spectra of the ternary crystals ZnSe:$Fe^{2+}$ $Zn_{1-x}Mg_xSe$:$Fe^{2+}$ with different solid solution concentration (activator concentration ~$5 \cdot 10^{18}$ cm$^{-3}$).

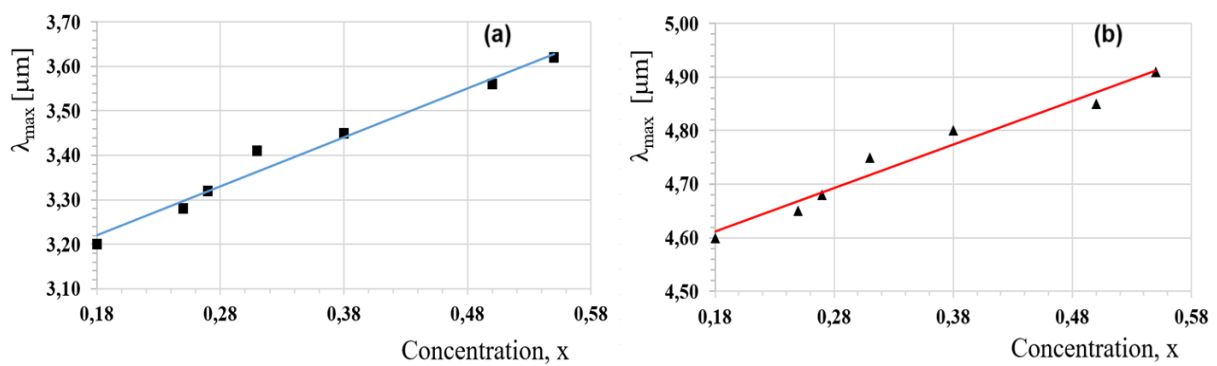

**Fig. 8.** Redshift of the maxima of $Fe^{2+}$ ions of absorption band (a) and emission band (b) for the ternary crystals $Zn_{1-x}Mg_xSe$:$Fe^{2+}$ with different solid solution concentration.

Analogous to the absorption spectrum of the binary ZnSe:$Fe^{2+}$ crystals, the absorption spectra of the solid solution crystals $Zn_{1-x}Mg_xSe$:$Fe^{2+}$ also split into three characteristic bands. When the solid solution concentration changes, a redshift occurs for each of the individual bands and the spectrum as a whole. Fig. 9 shows the shifts in the three Jan-Teller absorption sub-bands ($\lambda_1 < \lambda_2 < \lambda_3$) and the changes in their widths when the solid solution concentration changes from $x = 0$ to $x = 0.38$. These three components indicated correspond to the deconvolution of the full absorption spectrum into three main Jan-Teller sub-bands (see Fig. 7). Notably, with the same change in the concentration of the solid solution, the longer-wave components shift more strongly than the shorter-wave components do; thus, the following relationship always holds:

$$\Delta\lambda_1(x) < \Delta\lambda_2(x) < \Delta\lambda_3(x). \tag{3}$$

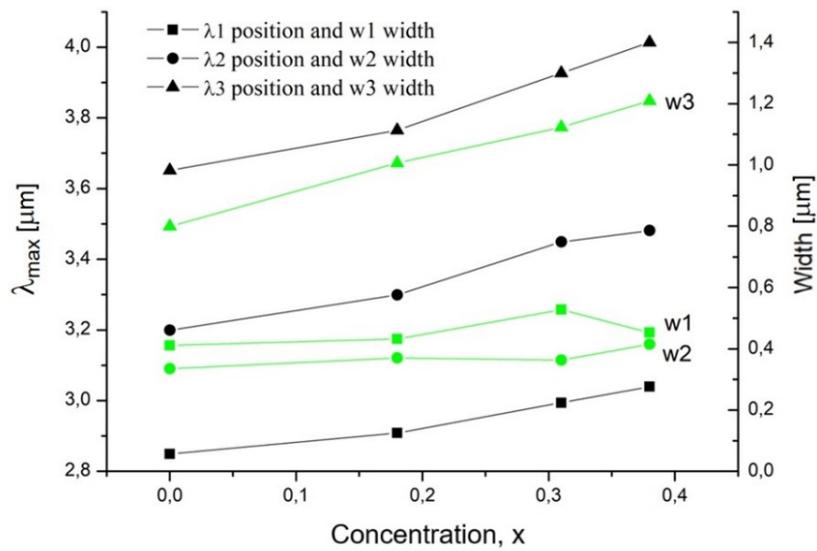

**Fig. 9.** The redshift of the spectra maximum position and width of the Jan-Teller absorption sub-bands of $Fe^{2+}$ ions in the ternary crystals $Zn_{1-x}Mg_xSe$:$Fe^{2+}$ at room temperature.

The longer-wavelength spectral absorption sub-bands shift much more strongly with increasing solid solution concentration than the short-wavelength optical sub-bands do. A kind of optical band segregation effect occurs. As a result of this effect, the extremal components (with the longest and shortest wavelengths) of the optical transitions move apart as the concentration of the solid solution increases. As a consequence, the widths $(w_1, w_2, w_3)$ of sub-band components and the total absorption band as a whole broadens. This phenomenon is clearly observed from a comparison of the optical transmission spectra, which for crystals with a higher solid solution concentration turn out to be less sharp.

**Conclusions**

Doped semiconductor crystals of binary ZnSe:$Fe^{2+}$ and ternary $Zn_{1-x}Mg_xSe$:$Fe^{2+}$ compounds were grown using the high-pressure Bridgman method and comprehensively studied. The correlations between variations in solid-solution composition, structural and electronic properties, and the positions of the $Fe^{2+}$ ion absorption and emission maxima were analyzed. An explanation was provided for the observed redshift of the mid-IR absorption and emission bands in the spectra of transition-metal ions in $Zn_{1-x}Mg_xSe$:$Fe^{2+}$ crystals with increasing solid-solution concentration. The redshifts of the total absorption and emission maxima were approximately 100 nm and 80 nm, respectively, for every 10% increase in solid-solution concentration. The phenomenon of a differentiated redshift of the Jahn–Teller components (sub-bands) of the total absorption spectrum was also discussed. The obtained results can be used to predict the lasing range of $Zn_{1-x}Mg_xSe:Fe^{2+}$ active materials across the entire range of possible Mg concentrations.

**Acknowledgements**

This work was supported by the National Research Foundation of Ukraine (NRFU), grant number 2025.06/0006.